\documentclass[a4paper,11pt]{article}
\pdfoutput=1 
\usepackage[utf8]{inputenc}
\usepackage{jheppubmod}

\usepackage{graphicx,xcolor,amsmath,amssymb,fancyhdr,siunitx,mathtools}
\usepackage[numbers]{natbib}
\usepackage[titletoc]{appendix}
\numberwithin{equation}{section}

\usepackage{cleveref}

\def\dslash{\not{\hbox{\kern-2pt $\partial$}}}
\def\Dslash{\not{\hbox{\kern-4pt $D$}}}
\def\Oslash{\not{\hbox{\kern-4pt $O$}}}
\def\Aslash{\not{\hbox{\kern-4pt $A$}}}
\def\partialslash{\not{\hbox{\kern-4pt $\partial$}}}
\def\Qslash{\not{\hbox{\kern-4pt $Q$}}}
\def\pslash{\not{\hbox{\kern-2.3pt $p$}}}
\def\kslash{\not{\hbox{\kern-2.3pt $k$}}}
\def\qslash{\not{\hbox{\kern-2.3pt $q$}}}
\def\svslash{\not{\hbox{\kern-2.3pt $sv$}}}
\usepackage{verbatim}

\newcommand{\md}[1]{\textcolor{blue}{[MD: #1]}}

\title{On the Possibility of Demonstrating Confinement in Non-Supersymmetric Theories by Deforming Confining Supersymmetric Theories}

\author{Michael Dine,}

\emailAdd{mdine@ucsc.edu}

\affiliation{
    Santa Cruz Institute for Particle Physics and Department of Physics,\\University of California, Santa Cruz,\\
    Santa Cruz, CA, USA
}

\abstract{We recall that supersymmetric QCD with $N$ colors and $N_f \ge 0$ massive quarks confines.  We explain that this remains the case for small soft breakings and
small supersymmetric masses for any quarks.  These results
continue to hold at large $N$.  We then consider these theories with large soft breakings -- real
QCD --describing the problem in terms of the direct computation of the linear potential between a heavy quark and antiquark.  We provide
naive arguments that confinement persists, and then give a more general treatment, noting some features in common with $CP^N$ models .  We isolate one potential obstacle and discuss its plausibility.}

\begin{document}
\maketitle

\section{Introduction:   Arguing for Confinement in Non-Supersymmetric Theories Obtained by Deformation of Confining Supersymmetric Theories}

Confinement and the existence of a mass gap are readily established features of many $N=1$ supersymmetric gauge theories in four dimensions.  This is the case, in particular, for gauge theories without matter fields and gauge theories with only massive matter.  Exploiting tools such as holomorphy, it is possible to study many aspects of the problem, such as the role of monopole condensation\cite{seibergwitten1,seibergwitten2}.  It is natural to ask whether these results can be extended
to demonstrate confinement in non-supersymmetric theories.  One
could start, for example, by taking a supersymmetric gauge theory without matter, and then introducing a gaugino mass term.   Due to the mass gap, for small gaugino mass both the mass gap and confinement persist.  The question is:  is there a deconfining phase transition as $m_\lambda$ becomes large compared to the QCD scale?

The analogous problem can be addressed in $CP^N$ theories in two dimensions at large $N$\cite{wittencpn},  In this case, the non-supersymmetric theory (the theory without fermions) is confining, while
the supersymmetric theory is not.  As we will remark, even with large (but finite) masses for the fermions of the supersymmetric theory, the theory is not confining.
It will
be crucial for our discussion of QCD and the $CP^N$ models that confinement is a phenomenon of the far infrared theory.
This may be obvious, but it is worth pausing briefly to understand this point.

\subsection{Confinement as a Phenomenon of the Far Infrared Theory and its potential implications}

We consider in this subsection the four dimensional situation.  The application to two dimensions is immediate.  As we will shortly explain, the theory with a small gaugino mass is
confining.

This domination of the far infrared can be seen by considering the corresponding path integral expression for the expectation value of a large Wilson loop, $W$:
\begin{equation}
\langle W \rangle =\int [dA^\mu(k^0,\vec k)]_{\vert \vec k \vert < \Lambda_w, k^0 \ll \vert \vec k} W \int [dA^\mu(k^0,\vec k)]_{\vert \vec k \vert > \Lambda_w, k^0 > \vert \vec k\vert} e^{-S}.
\end{equation}
For a large Wilson loop, the second integration gives a suppressed contribution at small $\vert \vec q \vert$ (large $R$).  it is only the integration region of small momentum which can lead to a linear potential
at large distances.  Writing this expression as:
\begin{equation}
\langle W \rangle =\int [dA^\mu(k^0,\vec k)]_{\vert \vec k \vert < \Lambda_w, k^0 \ll \vert \vec k} W e^{-S_{\Lambda_w}}
\end{equation}
it follows that the theory with action $S_{\lambda_w}$ is confining, i.e. a pure gauge theory with the corresponding action is confining.

\subsection{Features of the Far Infrared Theory}

The infrared theory is different in the $CP^N$ theories with and without massless fermions, leading to drastically
different behaviors.  The question in four dimensions is:  could this possibility be realized in reverse, with confinement in the supersymmetric limit
and non-confinement in Real QCD.  In four dimensional gauge theories, we don't have a simple (continuum) picture of confinement, so the problem
is more challenging.  
If we deform the supersymmetric theories with small soft breaking terms, confinement will survive, as readily follows from the existence of a mass gap.
But for large soft breakings, i.e. to describe real QCD, more is required to establish confinement.  In \cite{murayamaconfinement}, arguments were  put forward that if one considered
theories with anomaly mediated supersymmetry breaking, there would be no phase transition.  These arguments relied on working only to linear order in soft breakings, and considering
the monopole condensate,
but as explained in \cite{dineyurealqcd}, non-linear effects (in the low energy dynamics) become important (and difficult to control) as we approach the QCD scale. So while the result is plausible, more is required to
make a compelling argument. 

But focussing on the heavy quark-antiquark potential (Wilson loop), we will argue that a deconfining transition is unlikely.
Framed this way, confinement follows from quite plausible  assumptions about the theory
in the far infrared, one of which, however, is particularly difficult to prove.  In large $N$, these assumptions reduce to properties which are generally assumed
to be true of the planar diagram expansion.          

\subsection{Outline of the Paper}

The rest
of this note is organized as follows. We briefly
review  confinement in supersymmetric theories in the next section.  In section \ref{cpn}, we then remark on confinement in the two-dimensional $CP^N$ models at large $N$, with and without supersymmetry, noting certain features which we expect to be relevant to four dimensions and some which are likely different. Turning to four dimensions, we give naive arguments for confinement and then a more general argument for large
$m_\lambda$ in section \ref{generalargument}, describing the most challenging loophole to close.  We explore the question of large $N$ theories in four dimensions with small and large soft breakings in section \ref{largen}.  Limiting consideration to CP conserving values of parameters reduces the problem of confinement to  standard assumptions about large $N$. In section \ref{lightquarks}, we extend
the analysis to theories with light quarks.  In section
\ref{etaprime}, we compare these assumptions with those governing the $\eta^\prime$ at large $N$\cite{wittencurrentalgebratheorems}.    In section \ref{thetadependence}, we remark on possible behaviors
with CP violating parameters. 
Our conclusions, that confinement likely holds in the theories with large soft breakings, along with the challenges to a proof, are presented in section \ref{conclusions}.

\section{Confinement in Supersymmetric Gauge Theories}
\label{susyconfinement}

Proving confinement in non-abelian gauge theories without supersymmetry remains one of the great challenges of theoretical physics.  What is remarkable is that, with $N=1$ supersymmetry, one can establish
confinement in a range of gauge theories, in a variety of ways.  These include:
\begin{enumerate}
\item  Following Seiberg and Witten\cite{seibergwitten1,seibergwitten2}, one can prove confinement by exhibiting monopole condensation in theories which can be obtained by deformation of $N=2$ theories. 
\item  Starting with $SU(N)$ supersymmetric QCD with $N-1$ massive flavors, one can perform a systematic computation, for small masses, of the spectrum and other features.   Using holomorphy,
one can study the theory in various regimes of quark mass, establishing, among other results, gaugino condensation and the existence of a mass gap in the $N_F = 0$ theory\cite{seibergholomorphy}.
\end{enumerate}

In the case of the $SU(N)$ theory with $N=2$ supersymmetry, before one adds a mass term for the adjoint field, $\phi$, there are $N$ points on on the moduli space of vacua where
monopoles become massless:
\begin{equation}
\langle u \rangle \equiv \langle {\rm Tr}~ \phi^2 \rangle = u_0 e^{2 \pi i \over N}.
\label{seibergwittenvacua}
\end{equation}
These transform into one another under a $Z_N$ symmetry.
Adding a large mass for the adjoint, the theory at low energies is supersymmetric $SU(N)$ QCD without matter, and these states correspond to the $N$ possible values of the gaugino condensate, $\langle \lambda \lambda \rangle \propto e^{2 \pi i \over N}$.
With a gaugino mass, the degeneracy is lifted, and all but one of the states 
are unstable.  For large enough gaugino mass, we would expect that all but the ground state are highly unstable.

So confinement and the existence of a mass gap are common in supersymmetric field theories.    Notably, they are features of certain large $N$ theories, such as
the $SU(N)$ supersymmetric gauge theory without matter, or with number of flavors, $N_f$, much less than $N$, providing additional tools to study non-supersymmetric theories.  

\section{Lessons from the $CP^N$ Models in Two Dimensions}
\label{cpn}



The $CP^N$ models at large $N$ in two dimensions\cite{wittencpn} are instructive for the case of four dimensional QCD.  Without supersymmetry, these theories are confining and gapped.  While confinement itself is much simpler in two than in four dimensions, other aspects of the problem, particularly the scaling of terms in a four dimensional effective action, have close parallels.
What is similar about the two theories is that confinement is a problem of the far infrared.  As a result, it is natural to focus on a Wilsonian action
with cutoff, $\Lambda_w$, well below the scale of the theory, $\Lambda$ (the dynamically generated scale in the two dimensional case, $\Lambda_{QCD}$,the QCD scale in the four dimensional case).  Confinement
is a feature of this very low energy effective theory.

For the non-supersymmetric $CP^N$ models, the microscopic degrees of freedom are a set of $N$ complex scalars on the $CP^N$ manifold; more conveniently, they are a set of scalar fields
along with a scalar lagrange multiplier and a vector field with no kinetic term at the classical level.   The scalars gain mass of order $\Lambda$, the dynamically generated scale of the
theory.  In the far infrared, the the scalar degrees of freedom are integrated out and the Wilsonian action is written in terms of
a gauge field, $A_\mu$, and its field strength $F_{\mu \nu}$.     
In the non-supersymmetric case, the low scale effective action can be computed in the loop expansion, with successive loops suppressed by by powers of $1/N$.  At
one loop, one has a series of operators of the form $F_{\mu \nu}^{2n}$, with coefficients scaling as $N \Lambda^{-4n+2}$.  But setting aside the explicit one loop
computation, we can understand this scaling by the requirement that if one perform a one loop computation with the effective action, it yields corrections to the effective action smaller than, or at least not parameterically larger than, the action itself.  Here we can take the ultraviolet cutoff to be (at most) of order $\Lambda$.  So, for example,
writing the coefficient of $F_{\mu \nu}^4$ as ${1 \over M^6}$, the one loop correction to the $F_{\mu \nu}^2$ coupling behaves as
\begin{equation}
N^{0}{1 \over M^{6}}\int^\Lambda d^2 k   {k^4 \over k^2} \sim N^0 {\Lambda^4 \over M^6}.
\end{equation}
This is (up to a factor of $N$) of order the leading term in the action, provided $M \sim \Lambda$.  We can place a sharper bound if we insist on a
particular behavior of the $1/N$ expansion; the result, needless to say, is in agreement with the explicit computation.

The hierarchy of operators is also consistent with confinement.  $F^2$ is the most relevant operator of the low energy theory
and the $F^2$ term in the action makes the dominant contribution to the potential at large distances.  For $\Lambda_w \ll \Lambda$,  the coefficient of $F^2$
does not evolve with decreasing $\Lambda_w$.
It is crucial that the infrared theory is confining for any value of the coefficient of the $F_{\mu \nu}^2$ term, $1 \over e_0^2$.
Changing $e_0$, the coefficient of the most relevant operator in the infrared theory, just rescales lengths in the large distance theory.
In four dimensions, we will argue for similar behavior.

The theory exhibits non-trivial $\theta$ dependence\cite{wittencpn}.   We will consider $\theta$-dependence later in four dimensions, where our arguments for
confinement will hold on the CP conserving subspace of couplings.

The supersymmetric version of the theory microscopically includes fermionic partners of the scalar fields.  It is convenient to work with the theory in the form described in \cite{wittencpn}, in which form it is
relatively easy to explicitly break the supersymmetry.  In terms of the scalars, $n_i$ fermions $\psi^i$, bosonic auxiliary fields $\lambda, \sigma, \pi, A_\mu$ and
fermionic auxiliariy fields $\chi, \bar \chi$, the lagrangian is: 
\begin{equation}
{\cal L} = {N \over g^2} [n_i^* (\partial_\mu - i A_\mu - i \pi)^2n^i + \bar \psi_i (i \partialslash - \Aslash) \psi^i - \bar \psi_i m_\psi \psi^i - {1 \over 2}(\sigma^2 + \pi^2) -\alpha m_\psi \sigma
\end{equation}
$$~~~~
- {1\over \sqrt{2}} \bar \psi^i(\sigma + i \pi \gamma_5) \psi^i
+ \lambda(n_i^* n^i -1) + \bar \chi n_i \psi^I + \bar \psi_i n^i \chi^*  ].
$$
Here we have included explicit supersymmetry violating terms, proportional to $m_\psi$.  The term linear in $\sigma$ is necessary for the renormalizability of the theory. Including $m_\psi$,
we can obtain, as in \cite{wittencpn}, the values of $\sigma,\pi$, and $\lambda$ at large $N$ by computing the effective action at one loop.  Setting $\pi = A_\mu = 0$, and differentiating the action with respect to $\lambda$ and $\sigma$ yields two equations:
\begin{equation}
N \int {d^2 k \over (2 \pi)^2} {1 \over k^2 + \lambda} = {N \over g^2};
\label{lambdaequation}
\end{equation}
\begin{equation}
N \int {d^2 k \over (2 \pi)^2} {\sqrt{2}({\sigma \over \sqrt{2}}  + m_\psi)\over k^2 + ({\sigma \over \sqrt{2}} + m_\psi)^2} = N( {\sigma\over g^2}) .
\label{sigmaequation}x
\end{equation}
Here we have noted that while in general there is a different renormalization of the fermion and boson kinetic terms, at large $N$ they are the same, so the same factor
of $1/g^2$ appears in both \ref{lambdaequation} and \ref{sigmaequation}.
The first equation is solved by 
\begin{equation}
\lambda \equiv \Lambda^2 =M^2 e^{-{4 \pi \over g^2}}.
\end{equation}
The second is solved, for small $m_\psi$, by $\sigma^2 = 2 \lambda$.   We can think of the scale, $\Lambda$, as a fixed physical scale, analogous to $\Lambda_{QCD}$, measuring $m_F = m_\psi + {\sigma \over \sqrt{2}}$ and other scales relative to this one.  For large $m_\psi$, ${\sigma \over \sqrt{2} } + m_\psi \sim m_\psi$.
We can understand this if we think of integrating out the fermion, with mass much greater than $\Lambda$,  In this case, the effects on the dynamics at low energies are
small .

At scales below the scale $\Lambda$, the system, in either limit, is described by the gauge field $A_\mu$ and the pseudoscalar field $\pi$.
These fields acquire non-trivial kinetic terms at one loop.  In the supersymmetric limit, the classical mass terms for $\sigma$ and $\pi$ are {\it cancelled}
by the one loop contributions.  But these fields still acquire mass, as explained in \cite{wittencpn}, as a consequence of kinetic mixing; the $\sigma$ field acquires the same mass:
\begin{equation}
M^2 = \Lambda^2 e^{-{4 \pi \over g(\Lambda)^2}}
\end{equation}
As a result, there is no linear potential between massive charged objects.  Indeed, in the infrared,  not only are there no particle excitations, but there are no gauge degrees of freedom.

When we include supersymmetry breaking,$m_F$, the mixing persists, but it is suppressed by $1 \over m_F$ for large $m_F$.  As a result, there is a range of distance for which the potential
grows linearly, but for sufficiently large $R$, the potential falls off exponentially.  Confinement is lost for any finite $m_F$.
This can be understood in the language of the effective theory.  For sufficiently small $\Lambda_w$, the theory does not even contain the gauge degrees of freedom.  

In the non-supersymmetric theory  ($m_F \rightarrow \infty$), the low energy theory is the $U(1)$ gauge theory.  We can think of a Wilsonian effective action, obtained by integrating out physics at scales above $\Lambda_w$.  $\Lambda_w$ can be arbitrarily small; we still find a linear potential.  This potential arises from the most relevant operator, $F_{\mu \nu}^2$, with coefficient
$-{1 \over 4 e_0^2}$.  This is a reflection of the fact that the linear potential is a phenomena of the far infrared in this theory -- arbitrarily far.  Less relevant operators yield
contributions to the potential suppressed by powers of $(\Lambda R)^{-1}$.  The coupling $e_0^2$ is proportional to $\Lambda=\Lambda(\Lambda_w)$, which we can now think of as a
coupling in the Wilsonian theory.  For small $\Lambda_w$, examining equation \ref{lambdaequation}, we see that $\Lambda$ goes to a constant as $\Lambda_w \rightarrow 0$
as $({\Lambda_w \over \Lambda})^2$.
As a result, the coefficient of the separation $R$ in the linear potential tends to a constant.x

In four dimensional gauge theory, we will see that confinement is a feature of the {\it supersymmetric} limit,  and that it survives, at least for small
supersymmetry breaking.  Provided that the far infrared theory exhibits certain {\it universality} properties, confinement follows even for large supersymmetry breaking ($m_\lambda \rightarrow \infty$).

It is worth noting that confinement occurs in a regime where there are no physical excitations of the system, so Greens functions of gauge-invariant local
operators fall to zero exponentially rapidly with growing separation.  Such correlators can be represented by contact terms in the low energy theory,; the value of these contact terms can be thought of as information which supplements the Wilsonian action.  
The Wilson loop is different; not being
a correlation function of a local operator, it does not satisfy a simple spectral relation.   Its large distance behavior can be computed from the Wilsonian action with arbitrarily small $\Lambda_w$. There are lessons here for four dimensions.   In particular, for small $m_\lambda$, the theory with $N_f=0$ is
gapped, confining, and has non-trivial $\theta$ dependence, implying a non-vanishing contact term in the $F\tilde F$ two-point function.  This is necessary to give the correct $\theta$-dependence, and should be
thought of as a feature which supplements the Wilsonian action.  Similarly, the large radius Wilson loop, corresponding to a charge which is a fraction of the charges of the $n_i$ and $\psi_i$ fields, obeys an area law, which can be computed from the small $\Lambda_w$ Wilsonian action.  The question is:  do these features change as we take $m_\lambda$ large?  What might change
in this case as the theory flows to small $\Lambda_w$?

It is also worth noting some parallels to the four dimensional situation.  In the confining regime in four dimensions, there is a monopole condensate.  For a heavy quark-antiquark system, the equation
\begin{equation}
\vec \nabla \cdot E = \rho
\end{equation}
implies an electric field between the quark and antiquark forming a flux tube with width of order $\Lambda_{QCD}$.  This geometry is effectively two dimensional, leading to a linear potential as in
the $CP^N$ model.  There are, again, no physical particle excitations at large distances.  The question is:  does the monopole condensate disappear at some critical value of the gluino mass $m_\lambda^0 = c \Lambda_{QCD}$, i.e. is there a
phase transition.  We will argue against such a possibility in the next section.  But the essence of the argument is simple.  Confinement is a phenomenon which can be studied arbitrarily far in the infrared.
For small $m_\lambda$, we know that this far infrared theory is confining.  Naively, this theory is a gauge theory with lagrangian $-{1 \over 4 g^2(\Lambda_w)}F_{\mu \nu}^2$.  The effect of changing $m_\lambda$
is just to change $g^2(\Lambda_w)$, leading to a confining theory with associated scale $\Lambda_{QCD}$.  If the system cannot be described in terms of these degrees of freedom, this may not hold.
In the language of monopole condensation, it is logically possible that the monopole condensate might disappear at $m_\lambda^0$, and the electric flux is no longer collimated for larger $m_\lambda$.

\section{An Argument for Confinement in Four Dimensions, and an Obstacle}
\label{generalargument}


Turning now to QCD,
to address confinement, rather than attacking the question of the fate of the magnetic monopole condensate at large $m_{\lambda}$, we focus on the heavy quark-antiquark potential.  Taking $R$ as the quark anti-quark separation (and $q$ the corresponding momentum transfer), for confinement we can study arbitrarily large $R$ (small $q$) for
general $m_\lambda$.  In particular, we can take $R \gg m_{\lambda}^{-1}$.

Restating this,
because confinement is a feature of the far infrared, it makes sense to consider a Wilsonian effective action with scale $\Lambda_w$.  Typically in thinking about this action in the continuum, one takes $\Lambda_w \gg \Lambda_{QCD}$, so that the action can be computed in a weak coupling approximation.  This would be appropriate to the problem of large $m_\lambda$, where integrating out the gaugino would yield a theory well described
in this way.  But the case of which we have knowledge is that of small $m_\lambda$, so we'll be interested in considering a range of $\Lambda_w$, including $\Lambda_w$ well
below the QCD scale.   The problem is then not a weak coupling one.

We might argue for confinement in real QCD by noting, first, that the supersymmetric theory is confining.  Because the theory is gapped, this remains
true in the presence of a small $m_\lambda$.  We can define a Wilsonian theory with scale $\Lambda_w$, integrating out physics about $\Lambda_w < m_\lambda$.
This pure gauge theory confines, with a characteristic scale $\Lambda_{QCD}$.  Now provided that the basic degrees of freedom in the far infrared are the same for all $m_\lambda$, we would
expect that all that changes as we change $m_\lambda$ is the scale $\Lambda_{QCD}$, and the theory is everywhere confining.  We elaborate on this argument now, but we note, from the start,
a loophole:
as a function of $m_\lambda$ the degrees of freedom of the theory might change, corresponding to the possibility of a phase transition.
In the magnetic description, for example, the disappearance of the magnetic monopole condensate at some finite value of $m_\lambda$ is a logical possibility.  This transition might be first or second order.


Consider, then, the softly broken $N=1$ gauge theory without matter fields as a function of $x={m_\lambda \over \Lambda_{QCD}}$.
We will restrict, first, to $\theta=0$.  More precisely we will take the $\theta$ parameter in the langrangian to be zero, $m_\lambda$ to be real, and the phase
of the gaugino condensate or of the field $u$ of equation \ref{seibergwittenvacua} to be zero.
For small values of $x$, $x < x_0$, for some $x_0$, we know that the low energy theory confines and is gapped.  Confinement is the result of some dynamics at scales well below the mass gap.  The non-supersymmetric $CP^N$ model illustrates this, as we have seen in section \ref{cpn}.  There it is the gauge field in the far infrared which is responsible for confinement.  This field, at low energies, does not create any of the particles of the theory, which are massive.

Integrating out physics above a scale $\Lambda_w$ slightly below the gluino mass, we are left with an effective theory with Wilsonian cutoff, $\Lambda_w$ slightly less than $m_\lambda$.  The pure gauge theory is confining for some ${\Lambda_w \over \Lambda_{QCD}} < y_0$, where $y_0$ is slightly less than $x_0$.    
Now provided that the
theory flows smoothly from larger $y$ to smaller, confinement is established.

For this to fail, the theory obtained by flow from large $\Lambda_w$ to small $\Lambda_w$, i.e. the theory obtained in the far infrared by integrating out a heavy
gluino must be different than that obtained  by integrating out a light gluino.  A possibility of this sort is suggested by the contrast between the supersymmetric and non-supersymmetric $CP^N$ models, which
have different degrees of freedom in the far infrared.  In the case of $CP^N$, though, this occurs for any value of the susy-breaking mass.  For QCD, we know that for small mass, the system is confined, so
we require a phase transition at some non-zero gaugino mass.  We will not be able to rule out this possibility, but we will see that it requires that the theory have some surprising properties.

We first give two somewhat naive arguments for confinement for large $m_\lambda$.  Each relies on the fact of confinement for small gluino mass.  The first starts by making the naive assumption that the dynamics of the theory
are well represented by the Feynman diagram expansion (suitably resummed).   Alternatively, one can ask under what circumstances there is a breakdown of the expansion.
In either case, a substantial enhancement of the individual contributions at large distances is then required to obtain a linear potential.  Such enhancements
would take the form of (powers of) logarithms of $q$.
For $q \ll \Lambda_w$ and $m_\lambda \sim \Lambda_w$, large logs only arise from diagrams with gluons.\footnote{Here we are working, as suggested above, with an effective theory where
the gluinos have been integrated out, giving contact terms in the action.}   So only gauge bosons play a role in confinement.  The confining potential, taken to be the
sum of these diagrams, is then a feature of the pure gauge theory, with Wilsonian cutoff $\Lambda_w$ and with a coupling $g(\Lambda_w)$ in the effective action.  The coefficient of the linear term in the potential, $k$, depends only on the scale parameter of the effective theory with small $\Lambda_w$, $\tilde \Lambda_{QCD}$, $k = a \tilde \Lambda_{QCD}^2$.  Here
\begin{equation}
\tilde \Lambda_{QCD} = \Lambda_w e^{-\int_{g_{ref}}^{g(\Lambda_w)}{dg^\prime \over \beta(g^\prime)}}.
\label{tildelambda}
\end{equation}
In this equation, $g_{ref}$ is some fixed reference value of the coupling.
  The form of this result will not change as we increase $\Lambda_w$ in the bosonic theory, as the effect is just a change in the overall scale.  So confinement survives at large $m_\lambda$.  One might question the reliability of the guidance provided by Feynman diagrams, but other known non-perturbative effects, such as instantons, are also dominated at large distances by the purely gluonic contributions.  The usual large $N$ picture, as we will discuss further below, assumes that features of QCD, at least at large $N$, are reflected
in the Feynman diagram expansion.  As for $\Lambda$ in the $CP^N$ model, $\tilde \Lambda$ tends to a constant as $\Lambda_w \rightarrow 0$, as follows from the fact of confinement in the small $m_\lambda$ theory.

We can reframe this argument, referring to a naive understanding of the operators of QCD, an understanding
which holds more rigorously at large $N$.  
For this argument for confinement,
we will make three assumptions:
\begin{enumerate}
\item  We can write an effective action, at large distances, in terms of gauge fields $A_\mu$ ($F_{\mu \nu}$).  This is plausible, and, at large $N$, a basic feature.
\item  Operators have roughly their naive dimensions.  In particular, there is a single marginal operator, $F_{\mu \nu}^2$, and associated dimensionless coupling, $g^2(\Lambda_w)$.
\item  The theory with small $m_\lambda$ and $\Lambda_w < m_\lambda$, defines the infrared sector of actual QCD, for some values of the parameters of the effective
lagrangian.  This is the case for large $m_\lambda$, where integrating out the gaugino generates a change in the marginal coupling and leaves over some irrelevant operators.
With our assumption above the situation is similar, in that the contributions of irrelevant operators to the potential at large distances will be suppressed
by powers of $(R \Lambda_w)^{-1}$.
\item  There are no zeros of the beta function.  As will see, and we can state more sharply in this formulation, we can actually rule out this possibility.
\end{enumerate}

With these assumptions, the theory in the far infrared has only one parameter, a mass or inverse length scale,  $\tilde\Lambda_{QCD}(\Lambda_w)$.  To define this, we proceed somewhat heuristically, writing an effective action
which governs very low scale phenomena ($\mu = m_\lambda,\Lambda_w$) as
\begin{equation}
{\cal L} = -{1 \over 4 g^2(\mu)} F_{\mu \nu}^2 + \rm \dots,
\end{equation}
and we have dropped higher dimension terms suppressed by powers of $\mu$.  Arguments of the type we gave for the $CP^N$ models bound these higher dimension terms
by powers of $1/\Lambda_w$ or ${1 \over \tilde \Lambda_{QCD}}$.
Indeed we can again define the scale $\tilde \Lambda_{QCD}$ as in equation \ref{tildelambda}.

We know that confinement occurs for small $m_\lambda$.  We know, as well, that the coefficient of the linear potential tends to a constant as $\Lambda_w \rightarrow 0$.
As a consequence, $\tilde \Lambda_{QCD}$ also tends to a constant as $\Lambda_w \rightarrow 0$.
This is similar to the behavior of the $CP^N$ theory.  Confinement, then, depends only on the length parameter $\tilde \Lambda_{QCD}$.  As we increase
$m_{\lambda}$, $\tilde \Lambda_{QCD}$ will change, but confinement will occur, simply associated with this different scale.


We can argue against an obstruction as we vary $g(\Lambda_w)$ by noting that a break in the flow would correspond to an infrared fixed point at some intermediate value of $\Lambda_w$.  If this were all there
is, then the theory in the small $m_\lambda$ region, starting with some $\Lambda_w$ would flow to this fixed point, which is inconsistent with the fact that it is gapped and confining.  So there must be at least two such points
(and in general an even number).  But now the theory would have an ultraviolet fixed point starting at small $\Lambda_w$, which is not consistent with the fact that the
theory with small $m_\lambda$ is asymptotically free.
At the level of planar Feynman diagrams, we can make diagrammatic versions of these statements, and we will do that in the next section.  

The argument for confinement can be stated more generally.
Confinement follows if:
\begin{enumerate}
\item  The theory at scales below $m_\lambda$, as we vary $m_\lambda$, has the same set of degrees of freedom.  This is essentially the assumption that there is no phase transition.  Note that this was
true of the $CP^N$ model with fermions where, at sufficiently low scale, there were no degrees of freedom at low energies, apart from the non-propagating
gauge field.  For QCD, there have been conjectures about the possibility of a Coulomb or conformal
phase.In the magnetic description, the disappearance of the monopole condensate at some finite value of $m_\lambda$ is a logical possibility\cite{seibergthetapi}.
Against this, we can only offer the (weak) argument that the very low energy theory, at weak coupling, is the same in the small and large $m_\lambda$ limits.
\item  In the low $\Lambda_w$ theory, there is a single, most relevant operator.  Correspondingly, for a given $\Lambda_w$, there is a single
parameter which characterizes the flow, $g(\Lambda_w)$.  $g(\Lambda_w)$ flows with $\Lambda_w$ according to the renormalization group.
\item  There are no fixed points in $g(\Lambda_w)$.
\end{enumerate}
If these statements are true, 
then the system flows from large $\Lambda_w$ to smaller $\Lambda_w$ where the system is known to be confining.  Equivalently, we can define
a parameter $\tilde \Lambda_{QCD}$, and again the confining parameter is independent of $\Lambda_w$.

In this more general framework, the possibility of a fixed point (or multiple fixed points) can again be ruled out by facts we know about the theory.  In particular, if there were a single fixed point, then starting with small $m_\lambda$ and
$\Lambda_w < m_\lambda$, the theory would flow at large distances to the fixed point.  But this is inconsistent with the fact of confinement in the small $m_\lambda$
theory.  If there were a pair of fixed points (more generally an even number of fixed points), we would also encounter inconsistency with known facts of the small $\Lambda_w$ 
theory.

This leaves the possibility of a phase transition, where the far infrared theory is described by different operators than the low energy theory.  We do not know how to rule out this possibility, but it is puzzling
for at least two reasons.
\begin{enumerate}
\item  To the extent we can think of the far infrared theory in terms of a sum of Feynman diagrams, these are the same in the small $m_\lambda$ ($\Lambda_w$) and large
$m_\lambda$ ($\Lambda_w$) limits.
\item  For small $\Lambda_w$, there would actually be two possible ``QCD"'s.   Information about the microscopic theory would be required to determine which was selected by nature.
\end{enumerate}

\subsection{A Path Integral Argument}

The argument for confinement can be cast in a path integral language. 
Because the problem is one of the far infrared, one can break the path integral into an integration above a scale, $\Lambda_w$, and below that scale.  For the Wilson loop, $W$, for example,
with radius large compared to $\Lambda_w^{-1}$ (corresponding to $q_0=0,\vert \vec q \vert \ll \Lambda_w$), we can treat the momenta of the $A^\mu$ degrees of freedom as Euclidean,
\begin{equation}
\langle W \rangle = \int [dA_\mu]_{\vert k_E \vert < \Lambda_w} W   \int [dA_\mu]_{\vert k_E \vert > \Lambda_w}[d\lambda]  e^{-S}
\end{equation}
We can summarize the second factor as $e^{-S_{eff}(A_\mu)}$.  If $S_{eff}$ can be written as a power series in $F_{\mu \nu}$, and is dominated by the $F_{\mu \nu}^2$ term,
confinement at large $m_\lambda$ follows.   This is basically a restatement of our arguments above in the case of the perturbation series.


\section{Confinement in the Large $N$ Expansion With Broken Supersymmetry}
\label{largen}

These arguments for confinement are particularly compelling in the large $N$ limit.  The main reason one might question the Feynman diagram argument is that Feynman
diagrams possibly might not capture the essence of the theory, but at large $N$, the planar diagrams {\it should} capture the theory's essential features.
Also at large $N$, the assumptions underlying the second argument are believed to hold:  the system is described, in particular, in terms of gluons and gluinos,
and the arguments used to bound the coefficients of operators in the effective action (and correspondingly the dimensions of these operators) should hold.

\section{Theories with $N_f$ Light Quarks At Large and Finite $N$}
\label{lightquarks}

We can extend these result to theories with $N_f$ light quarks.  Add to the action of this theory mass terms for the squarks and a mass for the gluino.  For small
values of the soft breaking parameters, one can find the ground state and calculate an effective action and the spectrum of the theory.  The theory is gapped and confined.  One can again
ask what happens when the soft breakings are increased, eventually to values larger than the QCD scale.   Once again, we can focus on the heavy quark-antiquark
potential.  At large distances, this is an infrared phenomenon, as before.
Taking the fermion masses to be non-zero but small, then as we did for the light gluino, we can consider the far infrared theory, and argue that this theory is the one we have
encountered in the theory with $N_f=0$.   Again this theory is
described by a single parameter and is necessarily confining.
At large $N$ with $N_f \ll N$, we can simply repeat the arguments of the $N_f =0$ theory.


\section{The QCD Description of the $\eta^\prime$ at Large $N$}
\label{etaprime}

While often unstated, the assumption that the underlying degrees of freedom at large $N$ in QCD in the far infrared are the gluon fields (and the dominance of planar diagrams) underlies the usual treatment of the $\eta^\prime$ at large $N$\cite{wittencurrentalgebratheorems}.  The analysis relies heavily on the anomaly equation,
\begin{equation}
\partial_\mu j^{\mu 5} = {2 N_f \over 16 \pi^2 } F \tilde F,
\end{equation}
or, in terms of the $\eta^\prime$,
\begin{equation}
f_\pi \partial^2 \eta^\prime = {2N_f \over 16 \pi^2}F \tilde F.
\end{equation}
From this, one can compute the $\eta^\prime$ mass in terms of the leading contribution in $1/N$ to the two point function of the $F\tilde F$ at distances of order a power of $N$ times $\Lambda_{QCD}^{-1}$
\cite{wittencurrentalgebratheorems}.  The result is expressed in terms of the correlator
\begin{equation}
U(k) = \langle (F(x) \tilde F(x) ~F(0)\tilde F(0) \rangle_k
\end{equation}
at zero (small, by a power of $N$) momentum. All of this is written explicitly in terms of the degrees of freedom arising from $A^\mu$.

Because the correlator   is evaluated at zero momentum, and because of the observational fact that the theory is gapped, it is, as in the
case of $CP^N$, controlled by a contact term
\begin{equation}
U(x) = K \delta(x).
\end{equation}
This contact term can be computed, for example, by computing the $\theta^2$ term in the action for a given $\Lambda_w$.  Because of the infrared sensitivity of the computation and because the system is gapped,
this term  is invariant
under changes of $\Lambda_w$.   Contributions to $U(k)$ from operators of higher dimension than $F_{\mu \nu}^2$ in the Wilsonian action are suppressed by powers of
${m_{\eta^\prime}^2 \over \Lambda^2} \sim N^{-1}$.


This is all quite parallel to the questions which arise for confinement:  one can study the deep infrared (small $\Lambda_w$) and work with the degrees of freedom of the
original continuum action.  The dominant term in the action is again the naive leading term.

\section{$\theta$ Dependence}
\label{thetadependence}

Once we allow a non-zero $\theta$, or equivalently treat $m_\lambda$ as complex, a richer phase structure is possible.
In particular, if we take $m_{\lambda}$ and $\langle \lambda \lambda\rangle$ as real, we can introduce a $\theta F \tilde F$ term, and our lagrangian now has two couplings of dimension four:
\begin{equation}
{\cal L} = - {1 \over 4 g^2_{eff}(\Lambda_w)} F_{\mu \nu}^2 + {\theta \over 16 \pi^2} F \tilde F. 
\end{equation}

Our discussion of the last section applies to the theory along the line $\theta =0$.
It established that confinement is likely in this space.  With non-zero $\theta$, a more intricate phase structure arises, and there are trajectories which
are sensitive to these phase changes. 
The supersymmetric gauge theory with small gaugino mass exhibits a branched structure, with phase transitions as one varies $\theta$.  Recall in particular that:
\begin{equation}
m_\lambda \langle \lambda \lambda \rangle \propto \vert m_\lambda \vert e^{i (\theta + {{2 \pi k} \over N})}.
\end{equation}
As a result, the theory undergoes phase transitions as $\theta$ varies.
 It is quite possible that there are regions in the space $(g_{eff}(\Lambda_w),\theta)$
with non-zero $\theta$ which evolve under the renormalization group to unconfined regions\cite{seibergthetapi}. 


\section{Conclusion:  Even With Large Soft Breakings, SUSY QCD is Almost Certainly Confining}
\label{conclusions}

From these rather simple considerations, it appears that, by deforming supersymmetric QCD with large gaugino and squark masses, one obtains
a theory which is gapped and confined.  
To summarize the basic argument:
\begin{enumerate}
\item  The linear confining potential of the supersymmetric theories survives with small soft breakings, due to the presence of a mass gap.
\item  The linear confining potential of the slightly broken supersymmetric theories can be computed at arbitrarily large distance.
\item  One can
work with a Wilsonian action with scale $\Lambda_w$ small compared to the soft breakings and large compared to $R^{-1}$.
\item  For small $\Lambda_w$ and large $R$, only the gluon fields contribute to the potential.
\item  The potential depends only on $\tilde \Lambda_{QCD}$, a parameter which depends in the usual renormalization group fashion on the value of $\Lambda_w$
and the most relevant coupling.  Increasing $m_\lambda$ just results in a rescaling of lengths.
\end{enumerate}
The last two points are basic features of the perturbative expansion.  For them to fail, we have argued, requires that the theory exhibit a phase transition at some finite $m_\lambda$.

For large $N$, the analysis is rigorous to the extent that large $N$ is defined
as a sum of an infinite set of planar diagrams, resummed in some way. 
Confinement in the theory with large breaking of supersymmetry, then, seems a likely consequence of the fact that the supersymmmetric theory is confining
and that confinement is a feature of the (arbitrarily) far infrared. 
  
\section*{Acknowledgments}
We thank P. Draper, G. Festuccia, N. Seiberg and E. Witten for conversations and critical comments.  We are particularly appreciative of the comments provided by members of the Simons Confinement Collaboration,
especially Ofer Aharony and Davide Gaiotto.
This work was supported in part by U.S. Department of Energy grant No. DE-FG02-04ER41286.

\appendices

\bibliography{real_qcd_from_susy_qcd}

\providecommand{\href}[2]{#2}\begingroup\raggedright\begin{thebibliography}{1}

\bibitem{seibergwitten1}
N.~Seiberg and E.~Witten, {\it {Electric - magnetic duality, monopole
  condensation, and confinement in N=2 supersymmetric Yang-Mills theory}},
  {\em Nucl. Phys. B} {\bf 426} (1994) 19--52,
  [\href{http://arxiv.org/abs/hep-th/9407087}{{\tt hep-th/9407087}}]. [Erratum:
  Nucl.Phys.B 430, 485--486 (1994)].

\bibitem{seibergwitten2}
N.~Seiberg and E.~Witten, {\it {Monopoles, duality and chiral symmetry breaking
  in N=2 supersymmetric QCD}},  {\em Nucl. Phys. B} {\bf 431} (1994) 484--550,
  [\href{http://arxiv.org/abs/hep-th/9408099}{{\tt hep-th/9408099}}].

\bibitem{wittencpn}
E.~Witten, {\it {Instantons, the Quark Model, and the 1/n Expansion}},  {\em
  Nucl. Phys. B} {\bf 149} (1979) 285--320.

\bibitem{murayamaconfinement}
C.~Cs\'aki, A.~Gomes, H.~Murayama, and O.~Telem, {\it {Demonstration of
  Confinement and Chiral Symmetry Breaking in SO(Nc) Gauge Theories}},  {\em
  Phys. Rev. Lett.} {\bf 127} (2021), no.~25 251602,
  [\href{http://arxiv.org/abs/2106.10288}{{\tt arXiv:2106.10288}}].

\bibitem{dineyurealqcd}
M.~Dine and Y.~Yu, {\it {Challenges to Obtaining Results for Real QCD from SUSY
  QCD}},  \href{http://arxiv.org/abs/2205.00115}{{\tt arXiv:2205.00115}}.

\bibitem{wittencurrentalgebratheorems}
E.~Witten, {\it {Current Algebra Theorems for the U(1) Goldstone Boson}},  {\em
  Nucl. Phys. B} {\bf 156} (1979) 269--283.

\bibitem{seibergholomorphy}
N.~Seiberg, {\it {The Power of holomorphy: Exact results in 4-D SUSY field
  theories}},  in {\em {Particles, Strings, and Cosmology (PASCOS 94)}},
  pp.~0357--369, 5, 1994.
\newblock \href{http://arxiv.org/abs/hep-th/9408013}{{\tt hep-th/9408013}}.

\bibitem{seibergthetapi}
D.~Gaiotto, A.~Kapustin, Z.~Komargodski, and N.~Seiberg, {\it {Theta, Time
  Reversal, and Temperature}},  {\em JHEP} {\bf 05} (2017) 091,
  [\href{http://arxiv.org/abs/1703.00501}{{\tt arXiv:1703.00501}}].

\end{thebibliography}\endgroup
\bibliographystyle{JHEP}

\end{document}